\documentclass[jkps,preprint,fleqn,showpacs,showkeys,10pt]{revtex4}
\usepackage{graphicx}
\usepackage{amssymb}
\usepackage{amsmath}
\usepackage{bm}

\usepackage{latexsym,amsfonts,braket,mathtools,comment}
\usepackage[dvipsnames]{xcolor}
\usepackage{epsfig}
\usepackage[all]{xy}
\begin{document}

\setcounter{page}{0}
\title[]{Almost certain loss from black holes:\\ critical comments on the black hole final-state proposal}
\author{Dong Jin \surname{Lee}}
\email{dj0626@kaist.ac.kr}
\affiliation{Department of Physics, KAIST, Daejeon 34141, Republic of Korea}
\author{Dong-han \surname{Yeom}}
\email{innocent.yeom@gmail.com}
\affiliation{Department of Physics Education, Pusan National University, Busan 46241, Republic of Korea}
\affiliation{Research Center for Dielectric and Advanced Matter Physics, Pusan National University, Busan 46241, Republic of Korea}

\begin{abstract}
In this paper, we critically revisit the Horowitz-Maldacena proposal and its generalization by Lloyd. In the original proposal, as well as in Lloyd's generalization, Hawking radiation involves a pair of maximally entangled quantum states in which the ingoing partner state and the collapsed matter form either a maximally entangled pair or a Schmidt decomposed random state near the singularity. We point out that the unitary matrix introduced in Lloyd's fidelity calculation depends on initial matter states; hence, his result on the high average fidelity may not represent an almost unitary evolution. In opposition to Lloyd's conclusion, when we do not include the state-dependent unitary matrix for the fidelity computation, we analytically and numerically confirm that information will almost certainly be lost because the fidelity will approach zero as the degrees of freedom increase. 
\end{abstract}



\maketitle

\section{Introduction}

The information loss paradox is a well-known, but unresolved, problem in modern physics \cite{Hawking:1976ra}. The issue revolves around how we can recover information from collapsed matter due to Hawking radiation \cite{Chen:2014jwq}. Despite this being a difficult task, it is accepted that if we assume several hypotheses, one may be able to obtain the unitary evolution of a black hole. A well-known approach in this regard is the \textit{Horowitz-Maldacena proposal}, which addresses the final state of black holes \cite{Horowitz:2003he}.

According to Horowitz and Maldacena, the following are assumed. First, Hawking radiation is a maximally entangled pair in which an antiparticle falls into and a particle exits a black hole. Second, the ingoing Hawking particles and collapsed matter particles are projected at the singularity. Third, when the ingoing Hawking particles and the collapsed matter particles are projected, the final state can be presented by a set of maximally entangled pairs. If we accept these three assumptions, information can escape perfectly by using the same mechanism as quantum teleportation.

The above three assumptions are too strong \cite{Hong:2008ga}. For example, Gottesman and Preskill \cite{Gottesman:2003up} noted that due to the interaction between infalling Hawking particles and collapsed matter, the maximal entanglement condition can break down (see also Ref.~\cite{Hwang:2017yxp}). As a result, a degree of information loss may occur. However, Lloyd \cite{Lloyd:2004wn} reported that such a loss of information is not particularly serious, and that based on reasonable fidelity, the original information can be recovered almost with certainty.

However, one important point is that Lloyd obtained the high average fidelity after multiplying a certain unitary matrix by the outgoing Hawking radiation. We would like to point out that the unitary operation used in the calculation depends on the state of the initial matter. Therefore, it makes subtle to say that the average fidelity can be a measure for unitarity.

Based on the above observation, we investigated the fidelity between collapsed matter and outgoing Hawking radiation once again. In contrast to Lloyd's estimation, in our research, the fidelity could be arbitrarily small as the number of degrees of freedom increased. This implies that as we naturally generalized the final state proposal, information had almost certainly been lost.

This paper is organized as follows. In Section~\ref{sec:rev}, we review the original version of the Horowitz-Maldacena proposal and Lloyd's generalization. In Section~\ref{sec:fidel}, we explain how fidelity can be a measure of unitarity and criticize Lloyd's calculation for general final states. In Section~\ref{sec:eff}, we propose the correct average fidelity when interactions exist between matter and the ingoing state. From this, we conclude that the Horowitz-Maldacena proposal indicates an almost certain loss of information. Finally, in Section~\ref{sec:con}, we summarize our results and discuss their possible applications and future research topics.

\begin{figure}
\begin{center}
\includegraphics[scale=0.33]{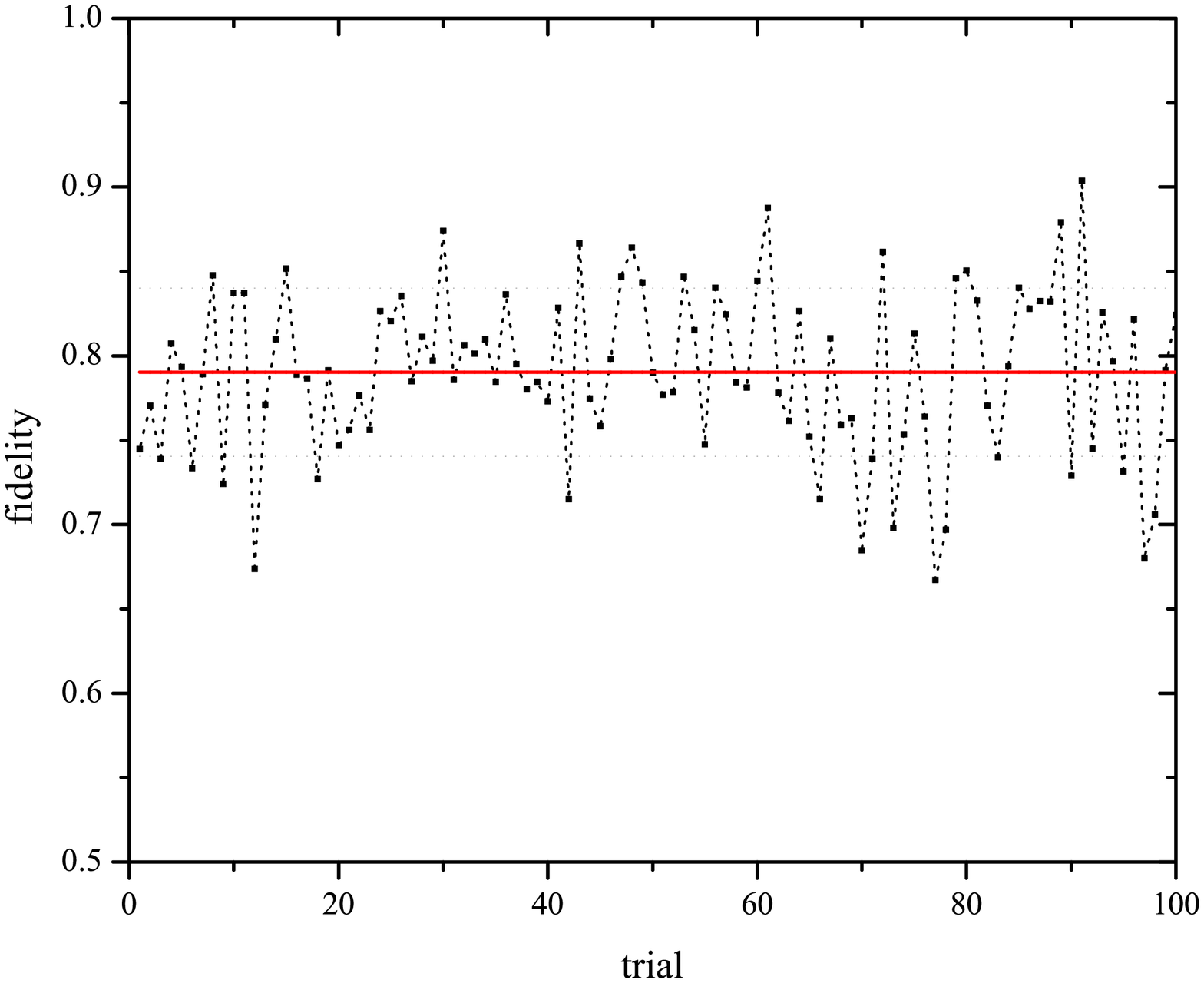}
\includegraphics[scale=0.33]{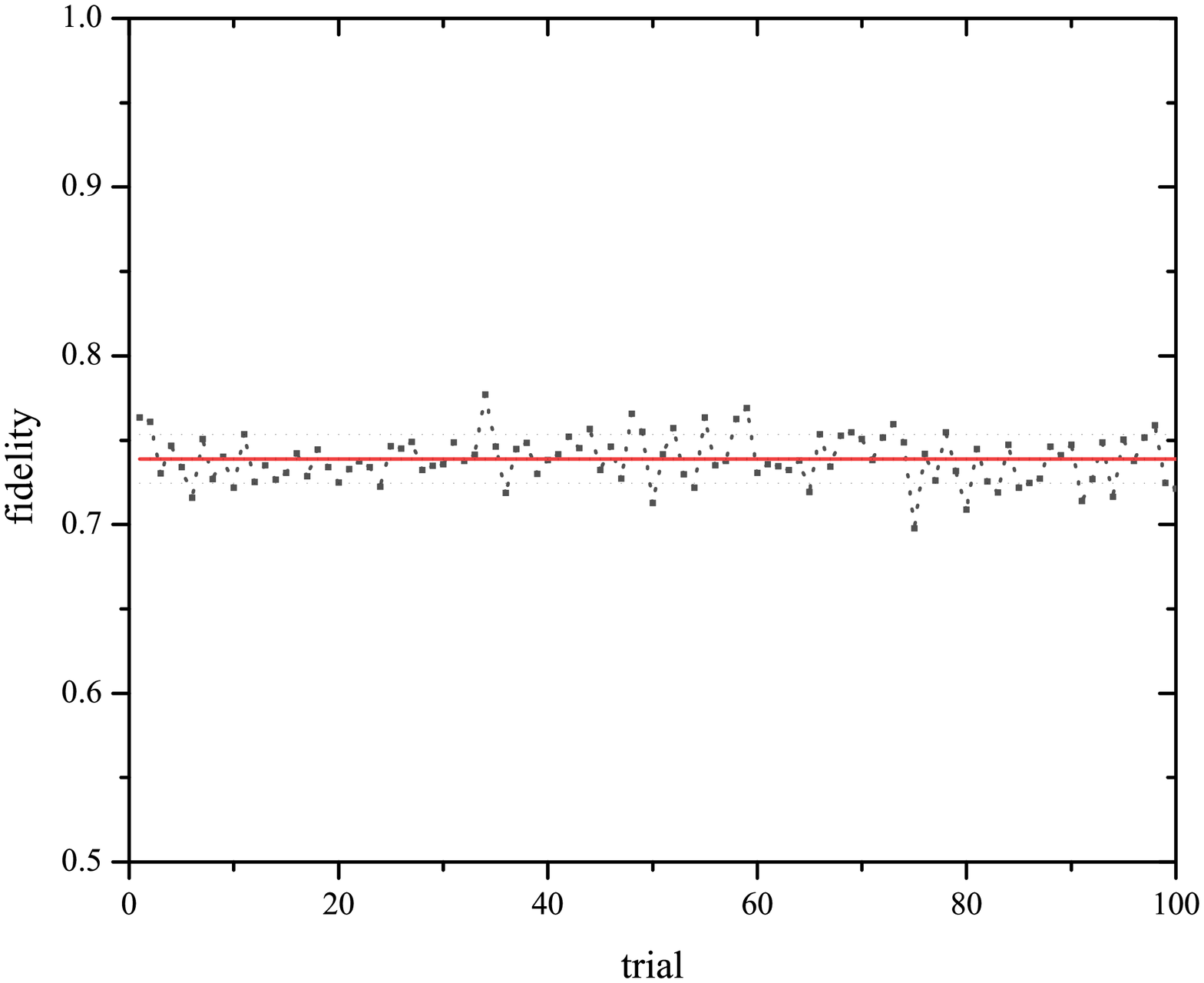}
\includegraphics[scale=0.33]{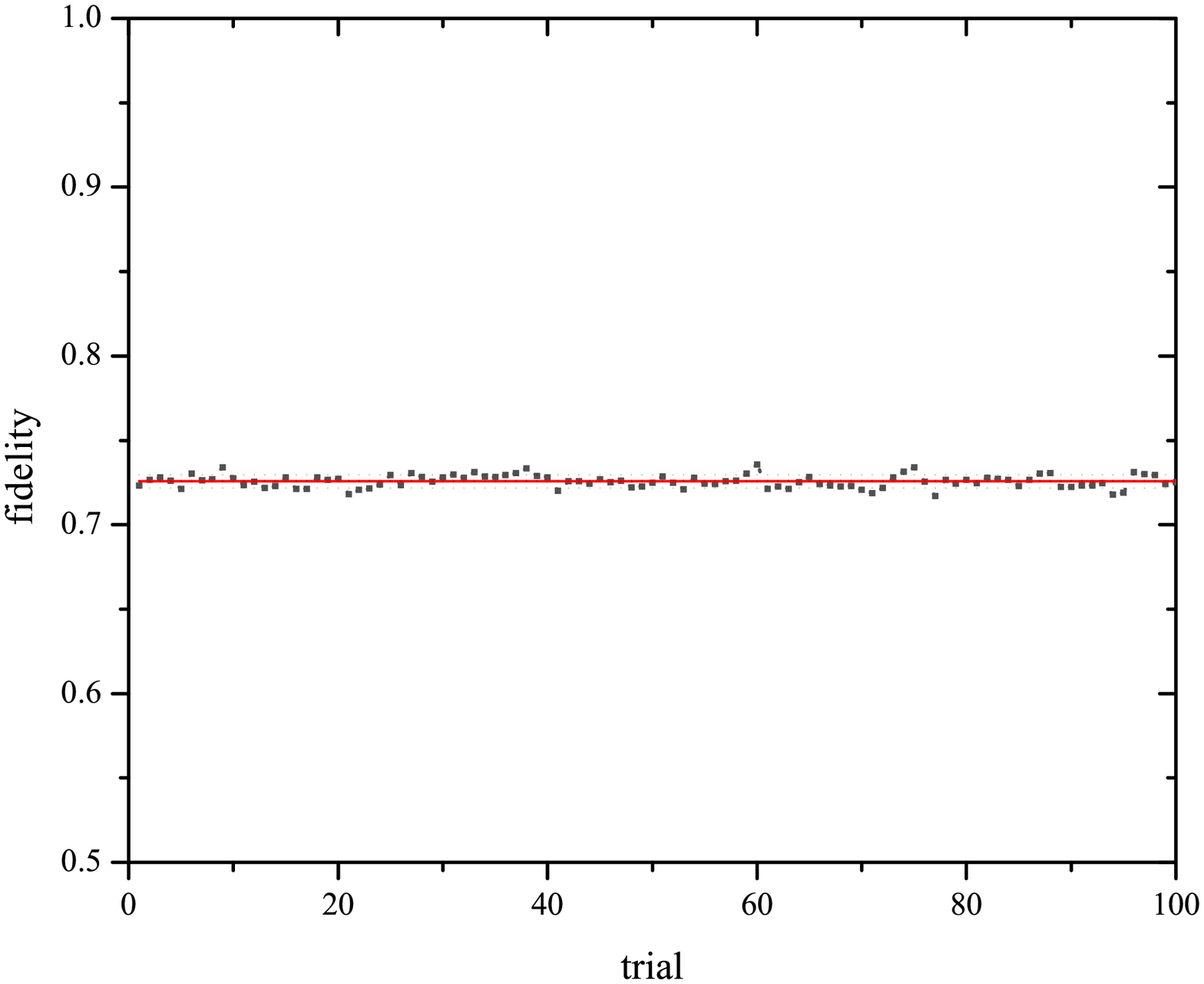}
\caption{Several fidelity results for $N=4$ (top), $N=16$ (middle), and $N=64$ (bottom) based on Lloyd's decomposition, where the red solid line indicates the average fidelity and the black dashed lines indicate the standard deviations. As the size increases, the standard deviation decreases and the average value approaches $(8/3\pi)^2 \approx 0.72$.}
\label{fig:1}
\end{center}
\end{figure}

\section{Review of the Horowitz-Maldacena proposal and the Lloyd generalization}\label{sec:rev}

Horowitz and Maldacena \cite{Horowitz:2003he} proposed that the final fixed state of a black hole at the singularity is able to resolve the black hole information problem. In their proposal, they considered the initial black hole system, which is a direct product of the matter state $\ket{\mu}_{\text{matter}}=\sum_{i=1}^N \mu_i \ket{i}_{\text{matter}}$ and the maximally entangled radiation states between incoming ($| j \rangle_{\mathrm{in}}$) and outgoing ($| j \rangle_{\mathrm{out}}$) particles, that is $\ket{\phi}_{\text{radiation}}=\sum_{j=1}^N \frac{1}{\sqrt{N}}\ket{j}_{\text{in}}\ket{j}_{\text{out}}$; in other words,
\begin{equation}\label{eq:1}
	\ket{\Psi}_{\text{initial}}=\ket{\mu}_{\text{matter}} \times \ket{\phi}_{\text{radiation}}=\sum_{i,j=1}^N \frac{\mu_i}{\sqrt{N}} \ket{i}_{\text{matter}}\ket{j}_{\text{in}}\ket{j}_{\text{out}}.
\end{equation}
The primary idea in this regard is that if the final state of the black hole is a maximally entangled state between matter and the ingoing states, after projecting the final state, the outgoing states preserve information about the initial state of the matter, which can be presented as follows:
\begin{eqnarray}\label{eq:2}
	\prescript{}{\text{BH}}{\bra{\Phi}}&=& \frac{1}{\sqrt{N}} \sum_{k=1}^N \prescript{}{\text{matter}}{\bra{k}}\prescript{}{\text{in}}{\bra{k}}\left(S\times I \right)  \;\;\;\;\;\;\;  (S: \text{unitary matrix}),\\
	\ket{\nu}_{\text{out}}&=& \prescript{}{\text{BH}}{\braket{\Phi|\Psi}}^{}_{\text{initial}}=\frac{S}{\sqrt{N}}\ket{\mu} \xrightarrow[]{\text{normalization}} S\ket{\mu}\label{eq:nor}.
\end{eqnarray}
Here, the reason why the outgoing state should be normalized in Eq.~(\ref{eq:nor}) is that the black hole final state is `post-selected' in the Horowitz-Maldacena proposal. When the final state of the black hole is fixed, a unique outgoing state can be observed from outside the black hole. Thus, the outgoing state should be renormalized, which, in turn, will preserve unitarity \cite{Horowitz:2003he,Gottesman:2003up}.

Soon after this idea was reported, Gottesman and Preskill \cite{Gottesman:2003up} pointed out that if interactions existed between matter and ingoing states, the unitarity of the entire process could be violated. Furthermore, they showed that the process of the final state projection is unitary if and only if the black hole final state is maximally entangled. Lloyd, however, showed that most of the random interactions at the black hole's final state preserved almost all initial information \cite{Lloyd:2004wn}. The black hole's final state following random interactions can be expressed in the following Schmidt form
\begin{equation}\label{eq:4}
\ket{\Phi}_{\text{BH}}=\sum_{k=1}^N \lambda_k \ket{k'}_{\text{matter}}\ket{k'}_{\text{in}}.
\end{equation}
The basis $\{\ket{k'}\}$ does not generally equal the $\{\ket{j}\}$ basis, and the distribution of the Schmidt coefficients $\lambda_k$ for random states is already well known \cite{Page:1993df}. By projecting Eq.~(\ref{eq:4}) into Eq.~(\ref{eq:1}), after writing the initial matter state in the $\{\ket{k'}\}$ basis, the final outgoing state becomes
\begin{equation} \label{eq:5}
\ket{\nu'}_{\text{out}}=\frac{1}{\sqrt{N}}\sum_{j,k=1}^N \lambda_k \mu_k \braket{k'|j} \ket{j}_{\text{out}}=\frac{1}{\sqrt{N}} \sum_{k=1}^N \lambda_k \mu_k \ket{k''}_{\text{out}},
\end{equation}
where $\ket{k''}_{\text{out}}= \sum_{j=1}^N \braket{k'|j} \ket{j}_{\text{out}}$. Lloyd introduced the unitary matrix $T'=\sum_{l,k=1}^N \ket{l'} \bra{k''}$ to match the basis of the initial matter state and that of the out-going state. After multiplying $T'$ by $\ket{\nu}_{\text{out}}$, the fidelity $f$ between the two states can be estimated as
\begin{equation} \label{eq:6}
f=|\langle \mu |T'| \nu' \rangle|^2 \approx N \left( \sum_j \mu_k^2 \lambda_k \right)^2 \approx \frac{1}{N} \left( \sum_k \lambda_k \right)^2\approx \left(\frac{8}{3\pi}\right)^2,
\end{equation} 
where $\mu_{k} \simeq 1/\sqrt{N}$ for random states. This estimation can easily be confirmed using numerical computations (Fig.~\ref{fig:1}, see Appendix). In this way, the fidelity can be accepted as remaining high despite the increase in the number of degrees of freedom $N$. Therefore, Lloyd claimed that the final state projection still conserved most of the information after the quantum error correction had been applied.

\section{How fidelity explains unitarity}\label{sec:fidel}

In this section, we explain how fidelity can be used as a measure of unitarity, especially for a process that is not exactly unitary. Then, one can conclude that multiplying by a unitary transformation $T'$ during the fidelity computation may not be proper.

For an unitary process, we can recover the initial state from the final state after having applied its inverse process. For example, if $\ket{\Psi}_{\text{out}}=U\ket{\Phi}_{\text{in}}$, then
\begin{equation}
f=\prescript{}{\text{in}}{\bra{\Phi}U^{-1}\ket{\Psi}_{\text{out}}}=1
\end{equation}
for any initial states $\ket{\Phi}_{\text{in}}$. Even though we do not know the process, if the fidelity between the final and the initial states becomes unity after having applied a proper unitary operation, we can conclude that the process can be described as the inverse of that unitary operation. In other words, if 
\begin{equation}
f=\prescript{}{\text{in}}{\bra{\Phi}U\ket{\Psi}_{\text{out}}}=1, 
\end{equation} 
for any initial states, then we can conclude that $\ket{\Psi}_{\text{out}}=U^{-1}\ket{\Phi}_{\text{in}}$.

One important point is that, in this case, the unitary operation must be independent of the initial state because the process itself should be consistent regardless of the initial input. For example, in the original version of the Horowitz-Maldacena proposal, the fidelity becomes unity after having multiplied by the inverse of $S$ (in Eq.~(\ref{eq:2})) to the out-going radiation state. This implies that the black hole evaporation process can be described by an unitary matrix $S$. This coincides with the well known procedure of the quantum error correction, which introduces a state-independent recovery operator to preserve the initial information \cite{Laflamme,Verlinde}.

The calculation of fidelity is useful to measure the preservation of information, especially, if the process is not perfectly unitary. If it is possible to make the fidelity close to unity using a proper unitary operation, one can conclude that the process approximately preserves information.

When the black hole final state is not maximally entangled, the final state projection process is not unitary because of the non-unitary renormalization. As in the Eq.~(\ref{eq:6}), Lloyd showed that the unitary operation $T'$ makes the average fidelity with random black hole final states close to one. From this result, he could conclude that the black hole final state projection was almost unitary for an arbitrary black hole final state as well as for arbitrary interactions between matter and the in-going radiation state.

However, we point out that the unitary operation $T'$ depends on the initial black hole state. Again, $T'$ can be explicitly presented as
\begin{eqnarray}
T'&=&\sum_{k=1}^N \ket{k'} \bra{k''}\\
  &=&\sum_{k=1}^N \sum_{j=1}^N \langle j|k'\rangle |k'\rangle \langle j|, \label{eq9}
\end{eqnarray}
where $\ket{k''}= \sum_{j=1}^N \braket{k'|j} \ket{j}$, $\{\ket{k'}\}$ is the Schmidt basis of the black hole final state, and $\{\ket{j}\}$ is the basis of the maximally entangled radiation state. Therefore, $T'$ depends on the Schmidt decomposition of the black hole final state. Unlike Horowitz-Maldacena's original proposal, we are considering a situation in which the black hole final state is determined by interactions between initial matter and in-going Hawking radiation. This implies that $T'$ should depend on the initial matter state. However, in order to consistently describe the quantum information transferring process, such a unitary transfer operation should not depend on the initial state. For instance, the unitary transfer operation included in the quantum teleportation protocol does not depend on the input states \cite{Bennett,Braunstein}.

Let us consider this argument for a simple case, i.e., when $N=2$. We can check the initial state dependence of $T'$ with the following set-up:
\begin{eqnarray}
\ket{\phi}_{\text{radiation}} &=& \frac{|0\rangle_{\text{in}}|0\rangle_{\text{out}} +|1\rangle_{\text{in}}|1\rangle_{\text{out}}}{\sqrt{2}}, \\
|\Phi\rangle_{\text{matter}, 1} &=& |+\rangle_{\text{matter}},\\
|\Phi\rangle_{\text{matter}, 2} &=& \frac{1}{2}|+\rangle_{\text{matter}} + i \frac{\sqrt{3}}{2}  |-\rangle_{\text{matter}}.
\end{eqnarray}
We can think of an interaction between the initial matter and in-going radiation as follows
\begin{eqnarray}
U_{\text{interaction}} 
= \begin{pmatrix} 1&0&0&0\\
0&\frac{1}{\sqrt{2}}&\frac{1}{\sqrt{3}}&\frac{1}{\sqrt{6}}\\
0&-\frac{1}{\sqrt{2}}&\frac{1}{\sqrt{3}}&\frac{1}{\sqrt{6}}\\
0&0&\frac{i}{\sqrt{3}}&-\frac{2i}{\sqrt{6}}
\end{pmatrix} \ ,
\end{eqnarray}
where the matrix is written with the $\{|+\rangle_{\text{matter}}|+\rangle_{\text{in}}, \; |+\rangle_{\text{matter}}|-\rangle_{\text{in}}, \; |-\rangle_{\text{matter}}|+\rangle_{\text{in}}, \; and |-\rangle_{\text{matter}}|-\rangle_{\text{in}}\}$ basis. Then, the final state for each initial matter state is given as follows:
\begin{eqnarray}
U_{\text{interaction}} |\Phi\rangle_{\text{matter}, 1}  |+\rangle_{\text{in}}&=&|+\rangle_{\text{matter}} |+\rangle_{\text{in}},\\
U_{\text{interaction}} |\Phi\rangle_{\text{matter}, 2}  |+\rangle_{\text{in}}&=&\frac{1}{2} \left( |+\rangle_{\text{matter}} |+\rangle_{\text{in}}+i|+\rangle_{\text{matter}} |-\rangle_{\text{in}}+i|-\rangle_{\text{matter}} |+\rangle_{\text{in}}-|-\rangle_{\text{matter}} |-\rangle_{\text{in}} \right) \\
&=& |\alpha\rangle_\text{matter} |\alpha\rangle_{\text{in}},
\end{eqnarray}
where $|\alpha\rangle \equiv \frac{|+\rangle + i|-\rangle}{\sqrt{2}}$ and $|\beta\rangle \equiv \frac{|+\rangle - i|-\rangle}{\sqrt{2}}$. From Eq.~(\ref{eq9}), we can obtain \textit{two different} unitary transfer operators for each initial matter state:
\begin{eqnarray}
T'_1 &=& |0\rangle \langle 0| + |1\rangle \langle 1|,\\
T'_2 &=& |0\rangle \langle 0| - |1\rangle \langle 1|.
\end{eqnarray}
Therefore, we can see that the unitary transfer operator generally depends on the initial matter state.

\section{The real effects of random interactions}\label{sec:eff}

Without the Schmidt decomposition, a random final state generally has the following form:
\begin{equation}\label{eq:9}
\ket{\Phi}_{\text{BH}}=\sum_{m,n=1}^N\lambda_{mn} \ket{n}_{\text{matter}} \ket{m}_{\text{in}}.
\end{equation}
In the above formula, $\ket{\Phi}_{\text{BH}}$ is expressed in a basis of maximally entangled Hawking radiation, and $\lambda_{mn}$ are not Schmidt coefficients but random complex numbers that satisfy
\begin{eqnarray}
\sum_{m,n=1}^N \lambda_{mn}\lambda_{mn}^*=1.
\end{eqnarray}
With this black hole final state, the final outgoing state becomes
\begin{eqnarray}
\ket{\nu}_{\text{out}}=\prescript{}{\text{BH}}{\braket{\Phi|\Psi}_{\text{initial}}} &=& \sum_{i,k,n,m}\frac{\mu_k}{\sqrt{N}}\lambda^*_{mn} \delta_{nk} \delta_{mi} \ket{i}_{\text{out}}\\
&=&\sum_{i,k}\frac{\lambda_{ik}^* \mu_k}{\sqrt{N}} \ket{i}_{\text{out}}.
\end{eqnarray}
The normalization factor is approximately given by
\begin{equation}
\braket{\nu|\nu}=\sum_{i,j,k} \frac{\lambda_{ik}^* \mu_k \lambda_{ij} \mu_j^*}{N} \approx\frac{1}{N}.
\end{equation}
Therefore, following normalization, the final state becomes
\begin{equation}
\ket{\nu}_{\text{out}}=\sqrt{N}\sum_{i,k}\lambda_{ik}^* \mu_k   \ket{i}_{\text{out}}.
\end{equation}

\begin{figure}
\begin{center}
\includegraphics[scale=0.33]{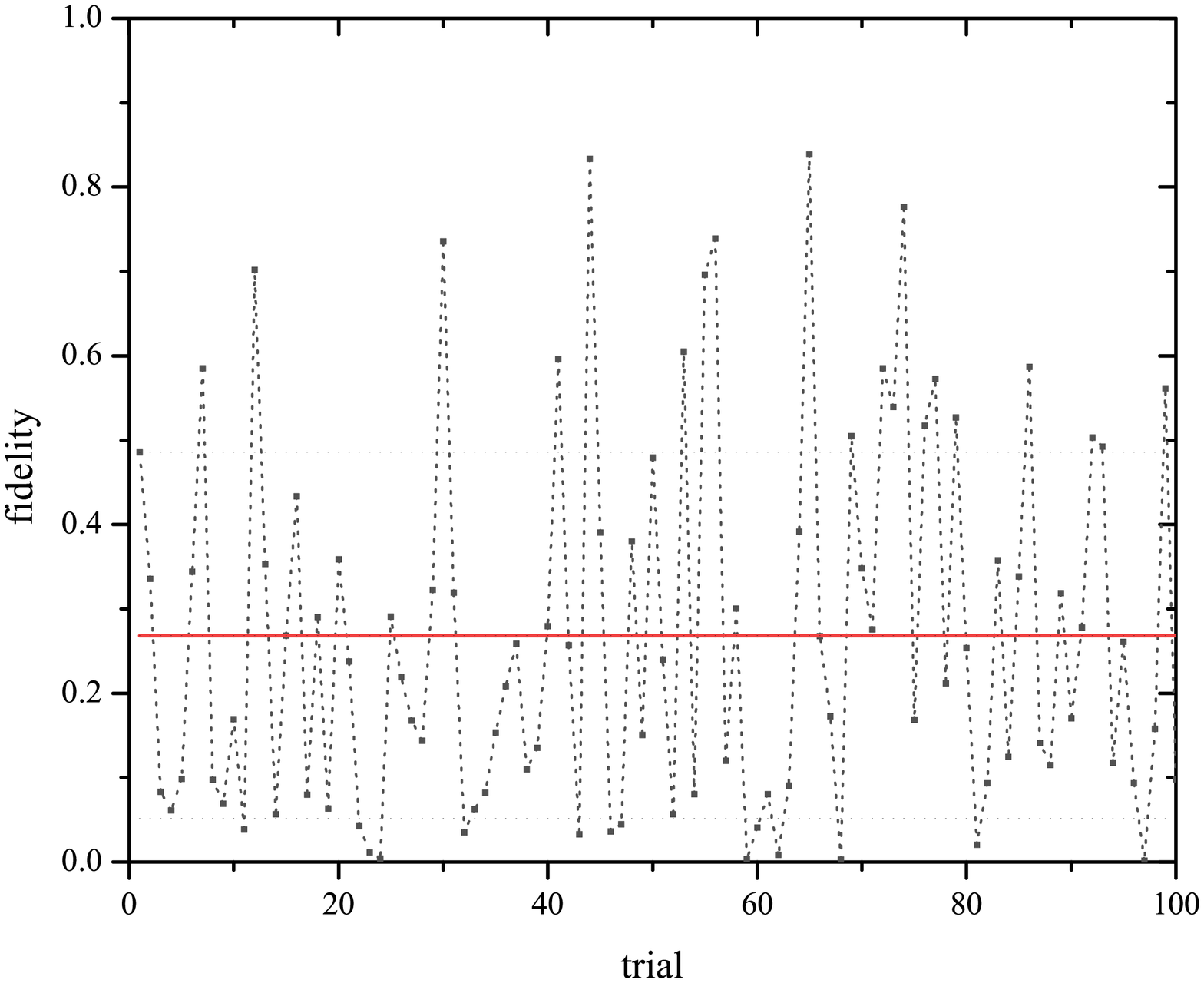}
\includegraphics[scale=0.33]{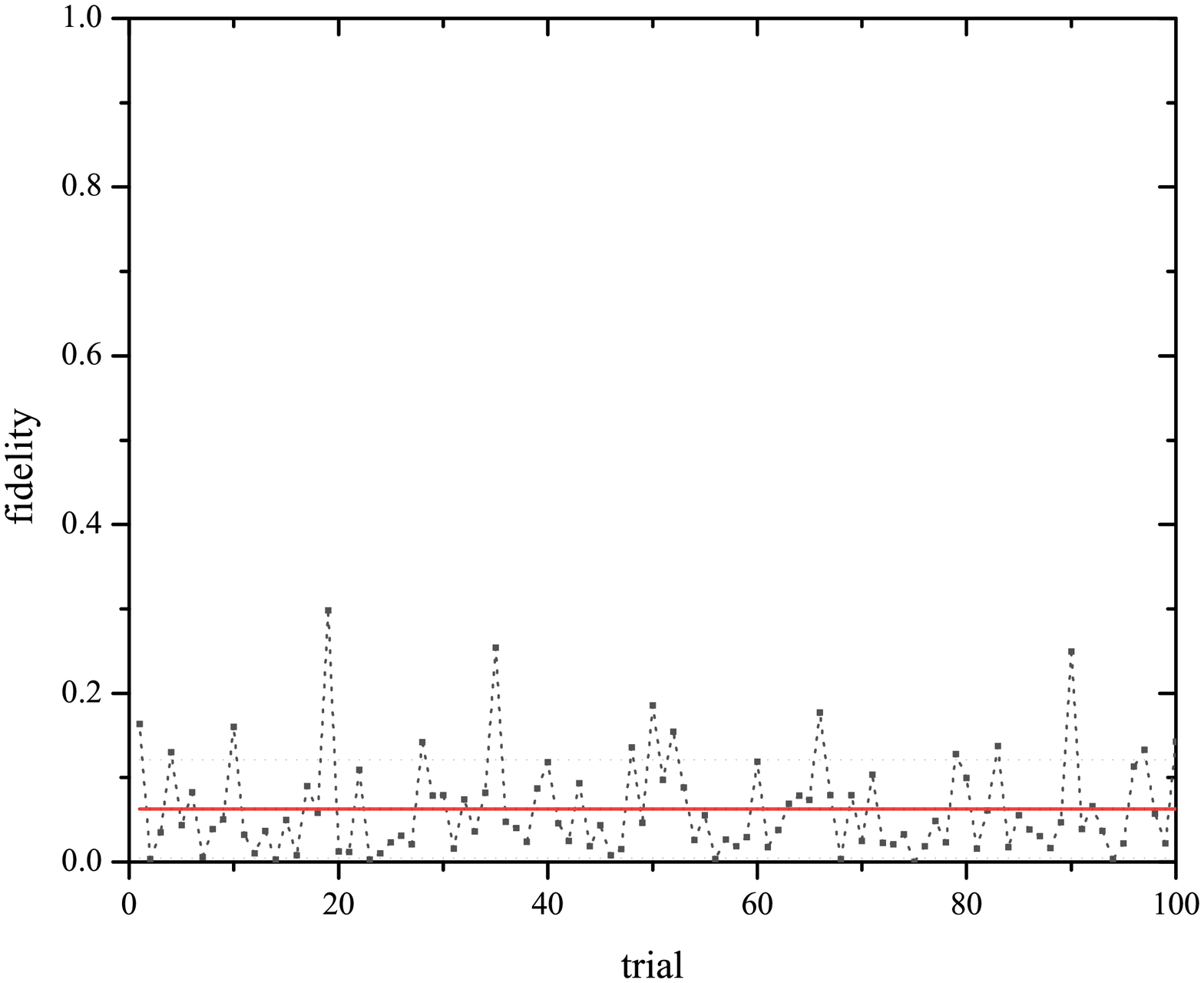}
\includegraphics[scale=0.33]{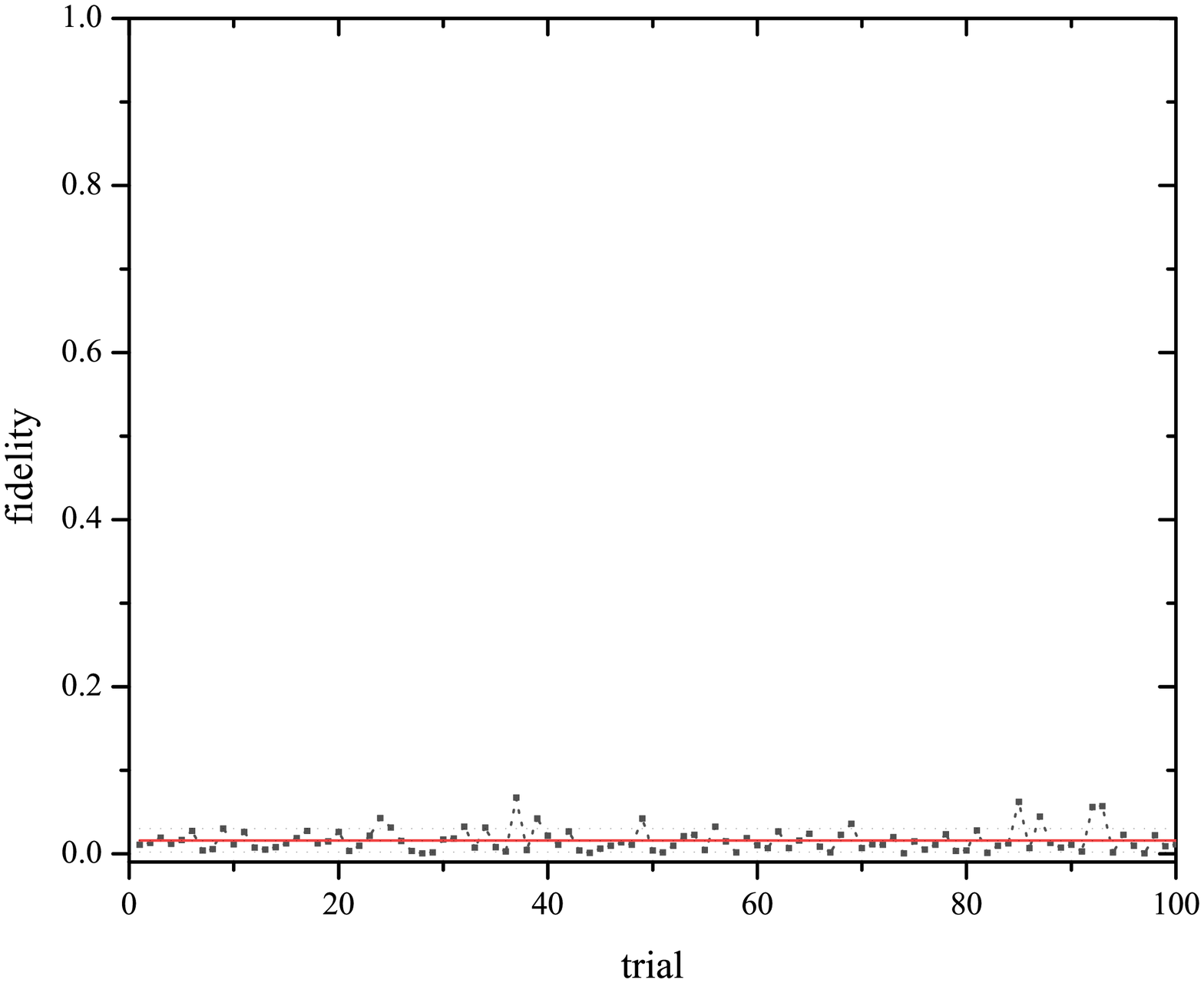}
\caption{Several trials of the fidelity for $N=4$ (top), $N=16$ (middle), and $N=64$ (bottom) cases assuming generic final states, where the red solid line indicates the average fidelity and the black dashed lines indicate the standard deviations.}
\label{fig:2}
\end{center}
\end{figure}

\begin{figure}
\begin{center}
\includegraphics[scale=0.33]{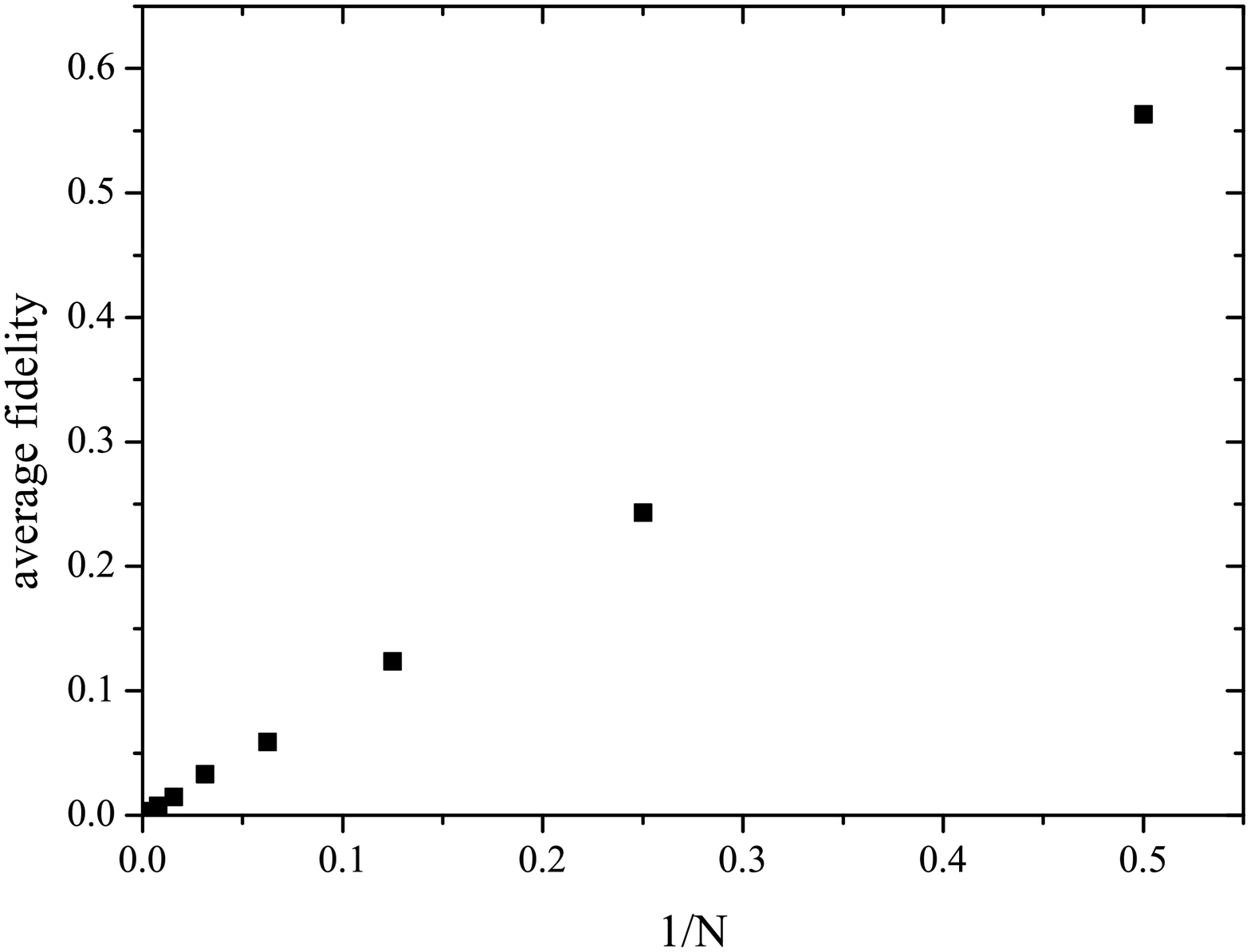}
\caption{Averaged fidelity as a function of $1/N$. This clearly shows its linear dependence.}
\label{fig:3}
\end{center}
\end{figure}

In order to examine the unitarity of the process, we should investigate whether the fidelity can remain high after multiplying a proper fixed unitary transfer operation $U$ by $\ket{\nu}_{\text{out}}$ for most of interactions, hence, for most black hole final states. However, the average of fidelities does not depend on the choice of $U$ because of the randomness of black hole final states. Without loss of generality, therefore, we can set $U=I$, and the average fidelity and its upper bound can now be computed as follows:
\begin{eqnarray}
f=|\langle\mu|\nu\rangle|^2 = N \left|\sum_{i,k}\lambda_{ik}\mu_k^* \mu_i \right|^2 \leq N\left(\sum_{i,k} \left| \lambda_{ik}\mu_k^* \mu_i \right| \right)^2 \approx \frac{1}{N},
\end{eqnarray}
where we have assumed $|\lambda_{ik}|\approx 1/N$ and $|\mu_i|\approx 1/\sqrt{N}$. This can be confirmed using numerical computations (Fig.~\ref{fig:2} and Fig.~\ref{fig:3}). 

Because the number of degrees of freedom of a black hole can be arbitrarily large, the fidelity can be arbitrarily small. Therefore, we can conclude that there is no guarantee of recovering information after having applied the quantum error correction; rather, it is more reasonable to think that the outgoing radiation state, obtained after the final state projection, does not conserve information on most interactions at the singularity. In other words, most states cannot be a black hole final state that conserves black hole information by outgoing radiation.

A worthwhile remark is that this conclusion is consistent with several results for quantum teleportation. We can consider the same problem using quantum teleporation because quantum teleportation with a fixed measurement is mathematically equivalent to the Horowitz-Maldacena proposal. The mean fidelity $\bar{f}$ attainable by using quantum teleportation with the Schmidt coefficients $\lambda_n$ is bounded by \cite{Banaszek}
\begin{eqnarray}
\bar{f} \leq \frac{1}{N+1} \left[1+ \left(\sum_n \lambda_n\right)^2 \right].
\end{eqnarray}
We can recover the result of Lloyd when the mean fidelity of quantum teleportation is maximized. However, it is maximized only if the state sender projects it to states that are maximally entangled via the optimal basis \cite{Banaszek}. In the case of a black hole, this corresponds to the condition that the Schmidt basis of the black hole final state be optimally matched to the basis of the maximally entangled radiation state, which Lloyd could have done by introducing $T'$. Therefore, for generic black hole final states, even though they are generally entangled states \cite{Page:1993df}, the mean fidelity is not maximized and converges to zero as the number of degrees of freedom of the black hole increases.

\section{Conclusion}\label{sec:con}

In this paper, we showed that for most random interactions in a black hole's final state, the outgoing radiation state loses most of the initial information present. This contrasts with Lloyd's result because the unitary matrix, which depends on the initial matter state, was introduced in his fidelity calculation. However, the process should be consistent regardless of the initial states; that is, the fidelity obtained after multiplying the initial-state-dependent unitary matrix cannot represent the unitarity of the process. With proper treatment of a random black hole's final state, the fidelity between the initial matter and the outgoing radiation states becomes zero as the number of degrees of freedom of the black hole increases. This implies that the interaction in the black hole's final state crucially violates the unitarity of the black hole. 

The discussion in Ref.~\cite{Lloyd:2013bza} indicates that the Horowitz-Maldacena proposal is useful for realizing black hole complementarity \cite{Susskind:1993if} when considering arguments about its inconsistency \cite{Yeom:2008qw,Almheiri:2012rt}. This sensitively relies on the assumption that unitarity is approximately well preserved despite relaxing the maximal entanglement condition for the final state. However, our discussion indicates that this is no longer guaranteed unless an unknown quantum gravitational principle that protects the quantum state, as is the case in the Schmidt form, exists.

In the original Horowitz-Maldacena proposal, the incoming Hawking particle is maximally entangled with its outgoing partner. Concurrently, the same incoming Hawking particle is maximally entangled with a quantum state of collapsed matter. Indeed, this is contradictory to the monogamy principle of entanglements (although one may further argue that the projection occurs at the singularity, where the principles of quantum mechanics may be violated). This is the essential point of the paradox among several assumptions related to natural laws \cite{Almheiri:2012rt,Hwang:2017yxp}. Accordingly, the only possible approach is to state the following: if Hawking radiation is maximally entangled with its partners and, at the same time, incoming partners are maximally entangled with star interiors, then their basis of quantum states must be presented differently.

Intuitively, Lloyd's ansatz state cannot be generic for the same reason. If no interaction between the star interior and incoming Hawking particles occurs, then, after the Schmidt decomposition, the maximally entangled pairs will be obtained. Then, their basis of quantum states must differ between maximally entangled pairs of Hawking radiation and final states in order to preserve the monogamy of entanglements. Due to this observation, unsurprisingly, the Schmidt decomposition cannot present the most general quantum states for the projection in terms of the basis of quantum states of Hawking radiation.

Our observations provide a deeper understanding of the tensions between several ideas about the information loss paradox. This may help to shed additional light on finding a new approach for ultimately resolving the paradox.

\begin{acknowledgments}
DY was supported by the National Research Foundation of Korea (Grant No.: 2018R1D1A1B07049126, 2021R1C1C1008622).
\end{acknowledgments}

\section*{Appendix: outline for numerical simulations}

To support the validity of our calculations, we performed numerical simulations for the fidelity between an initial matter state and the final outgoing radiation state. Two different simulations are presented in this paper. Fig.~\ref{fig:1} is the result of the simulation based on Eq.~(\ref{eq:4}) while Fig.~\ref{fig:2} is the result of the simulation based on Eq.~(\ref{eq:9}). A key point of both simulations was to generate random coefficients. In the former case, the random Schmidt coefficient $\lambda_n$ was needed; the latter required random coefficients $\lambda_{nm}$ and $\mu_k$, which satisfied $\sum_{n,m=1}^N \lambda_{nm}\lambda_{nm}^*=1$ and $\sum_{n=1}^N \mu_n\mu_n^*=1$, respectively. Accordingly, the fidelity for both cases was derived as follows. In Fig.~\ref{fig:1}, the fidelity is
\begin{eqnarray}
f \approx \frac{1}{N} \left( \sum_{n=1}^N \lambda_n \right)^2,
\end{eqnarray}
while in Fig.~\ref{fig:2} the fidelity is
\begin{eqnarray}
f \approx N \left|\sum_{n,m=1}^N \lambda_{nm}\mu^*_m\mu_n\right|^2.
\end{eqnarray}
To create random Schmidt coefficients, we first generated a $2N \times 2N$ random pure density matrix and partially traced out the matrix to obtain an $N \times N$ mixed density matrix. Then, the random Schmidt coefficients with the Hilbert-Schmidt measure were given by the eigenvalues of the mixed density matrix \cite{Zyczkowski}. For $\lambda_{nm}$ and $\mu_n$, we generated two random state vectors, one with $N^2$ dimensions and the other with $N$ dimensions. Next, we assigned each component of the vectors to $\lambda_{nm}$ and $\mu_n$.

\end{document}